\documentclass[letterpaper]{article} 
\usepackage[preprint]{aaai2027}
\usepackage[hyphens]{url}  
\usepackage{graphicx} 
\usepackage{natbib}  
\usepackage{caption} 
\usepackage{booktabs}
\usepackage{tabularx}
\usepackage{array}
\title{Who Can Evaluate AI Independently? A Framework for Analysing Evaluation Infrastructure}
\author{
    Sumaya Nur Adan\textsuperscript{\rm 1,2},
    Joanna Wiaterek\textsuperscript{\rm 2},
    Nourhan Abdelrhim\textsuperscript{\rm 3},
    Sydney Reis\textsuperscript{\rm 1},
    Robert F. Trager\textsuperscript{\rm 1},
    Benjamin Rosman\textsuperscript{\rm 3},
    Marina Jirotka\textsuperscript{\rm 1}
}
\affiliations{
    \textsuperscript{\rm 1}University of Oxford\\
    \textsuperscript{\rm 2}Centre for AI Security and Access\\
    \textsuperscript{\rm 3}University of the Witwatersrand
}

\begin{document}

\maketitle

\begin{abstract}
AI evaluation is increasingly central to AI safety and governance, yet taking part in an evaluation does not by itself confer authority over its design, its execution or the resources it depends on. This paper examines how infrastructure arrangements distribute that authority, and so shape who can evaluate AI systems independently. Drawing on Science and Technology Studies, political economy and applied evaluation standards, we introduce PAAI (Power, Access, Adequacy, Independence), a framework for examining who controls evaluation infrastructure, the terms on which it is available, whether it is adequate to the evaluation's purpose and what autonomy it supports. We apply PAAI to twelve African AI evaluation projects (2022--2025) and to the documented roles of all ten member institutes in the three published joint testing exercises of the international network of AI Safety Institutes. We identify three recurring configurations: extractive collaboration, dependent innovation and constrained independence. African actors often contribute data, language expertise and task design while external institutions retain authority over infrastructure and evaluation frameworks; design authority without infrastructure authority never yielded more than partial independence. The joint testing exercises show a similar division among mostly high-income members: the institutes that led each exercise also ran the models, while Kenya, the network's only African member, moved from annotation to task design with no documented model execution. Participation in evaluation does not by itself produce evaluative authority. For evaluation to serve as a credible governance mechanism, access must come with control over priorities, methods, publication and continuity.
\end{abstract}

\section{Introduction}\label{sec:intro}

When Kenya's AI Safety Institute participated in a 2025 international joint testing exercise on AI agents, Kenyan researchers constructed new tasks and tools on fraud, phishing and impersonation, translated and annotated test data in Kiswahili, and contributed to the cybersecurity analysis \citep{inais2025agenticreport}. Kenyan staff are credited co-authors of the resulting report \citep{seah2026agentic}. Singapore and the United Kingdom led the exercise and its principal testing and analysis activities, while no model execution by Kenya is documented. This division of work does not establish whether Kenya could have repeated the evaluation independently. It does, however, reveal a distinction between contributing substantive evaluation content and holding responsibility for the infrastructure and procedures through which it is tested. That distinction raises the question addressed in this paper: what forms of authority and access must accompany participation for evaluators to conduct and sustain assessments on their own terms?

This dynamic is not unique to Kenya. It reflects a structural tension at the heart of contemporary AI governance. By \emph{AI evaluation}, we mean the empirical assessment of AI models and systems to establish their safety properties, capabilities and risks, through methods such as red-teaming, capability benchmarking and domain-specific risk assessment \citep{ganguli2022red,liang2023helm,weidinger2023sociotechnical,shevlane2023model}. Independent evaluation of this kind is increasingly recognised as essential to AI governance \citep{bengio2025report}, yet the infrastructure that makes rigorous evaluation possible is controlled largely by the entities being evaluated or by institutions in a small number of well-resourced countries \citep{sastry2024computing,crawford2021atlas}. By \emph{evaluation infrastructure} we mean the material and institutional resources an evaluation depends on: the compute needed to run tests; access to the models being tested, whether only to their inputs and outputs or also to their internal workings \citep{casper2024blackbox}; test datasets and benchmarks; tools for running and scoring tests; and the funding, agreements and institutions that keep these available over time \citep{ojewale2025towards}. The result is not only a capacity gap but a \emph{control gap}. Actors may contribute to evaluation processes while lacking authority over evaluation priorities, methods, publication and continuity.

Building on standards for independent third-party evaluation \citep{stosz2025aef1} and work on external auditing of AI systems \citep{raji2022outsider,costanzachock2022who}, we define \emph{independent evaluation} as evaluation in which evaluators can exercise methodological and interpretive judgement and communicate their findings without undue influence from the developers whose systems are assessed or the organisations that provide the infrastructure for assessing them. This autonomy depends on two further conditions: the operational capability to execute and repeat tests, and the institutional continuity to sustain assessment over time. The three may support one another, but they are not interchangeable: an evaluator may publish independently while relying on precarious infrastructure, or possess substantial technical resources while facing restrictions on its methods or findings. We treat all three as necessary for independent evaluation.

Recent initiatives appear to address these gaps. Frontier AI developers offer API credits for safety research \citep{anthropic2024era}; cloud providers run academic programmes available in Africa \citep{googlecloud2024,microsoft2024azure}; and international collaborations promise distributed evaluation capacity \citep{inais2025multilingual}. Existing evaluation standards identify operating conditions that support the independence and transparency of particular assessments \citep{stosz2025aef1}, while research on AI infrastructure examines the distribution and governance of the resources on which AI development depends \citep{sastry2024computing}. Our concern is how these questions intersect: how do infrastructure arrangements distribute the authority, practical capabilities and continuity required for independent evaluation? We address this question through a framework that connects control over evaluation design and resources to the conditions under which evaluators can use them, and apply it to evaluation in Africa and to international joint testing.

\paragraph{Statement of Contributions.} We make four contributions. First, we introduce the PAAI framework (Power, Access, Adequacy, Independence), a framework designed specifically to characterise power arrangements in AI evaluation infrastructure, drawing on infrastructure studies, applied evaluation standards and the political economy of AI (Section~\ref{sec:framework}). Second, we apply PAAI through structured thematic analysis of twelve African evaluation projects, producing a cross-case mapping of evaluation infrastructure in Africa across three evaluation types (Section~\ref{sec:africa}). Third, we test whether these patterns reflect regional scarcity or the way shared evaluation is organised, by coding the role of every member institute in the three published joint testing exercises of the international network of AI Safety Institutes (Section~\ref{sec:jt}). Fourth, we derive three governance propositions and a set of design conditions that explain why expanding evaluation access can widen participation while entrenching dependency, and what would need to change (Section~\ref{sec:props}).

\section{Background and Related Work}

\subsection{AI Evaluation as a Governance Mechanism}

Beyond the methods noted above, AI evaluation includes bias auditing \citep{raji2020closing} and ethical risk assessment \citep{weidinger2021ethical}. A growing literature argues that such evidence is credible only if external evaluators can scrutinise models independently of developers \citep{raji2022outsider,anderljung2023towards,mokander2024auditing}. Yet \citet{raji2022outsider} demonstrate that the conditions enabling credible third-party oversight are unevenly distributed: model access, legal protections for auditors and institutional legitimacy all shape who can conduct consequential assessments. Developers decide who receives access to their models and in what form \citep{shevlane2022structured}; black-box access is insufficient for many rigorous audits \citep{casper2024blackbox}; and researchers who probe models risk losing access or facing legal threats, prompting calls for a safe harbour for evaluators \citep{longpre2024safe}. Field scans find that auditors lack shared standards, tools and protections, in an ecosystem that remains fragmented and concentrated in well-resourced contexts \citep{costanzachock2022who,ojewale2025towards,birhane2024auditing}. \citet{reuel2025who} map systematic gaps in who conducts AI social impact evaluations, finding that actors from the Global South are consistently underrepresented both as evaluators and as subjects of evaluation frameworks.

\subsection{Power Asymmetries in African AI Development}

Scholarship on AI in the Global South documents systematic asymmetries in development capacity and governance participation. \citet{abebe2021narratives} demonstrate that Western control over Africa's data and digital infrastructure shapes which AI applications are feasible, while \citet{birhane2020algorithmic} frames this as ``algorithmic colonization'', whereby systems developed elsewhere are deployed without addressing local needs. \citet{mohamed2020decolonial} use decolonial theory to show how such patterns extend historical relations of power, and \citet{okolo2023addressing} documents how global inequities in AI development capacity disadvantage actors in the Global South across the full AI pipeline. Fairness and evaluation concepts developed in Western settings often fail to transfer to other contexts \citep{sambasivan2021reimagining}. \citet{birhane2022power} examine how participatory AI initiatives reproduce similar asymmetries: formal inclusion does not translate into substantive power over design, deployment or evaluation decisions. \citet{png2022tensions} identifies a ``paradox of participation'', whereby Global South stakeholders are formally included in governance initiatives yet excluded from resource allocation and priority-setting. Much AI labour, including annotation, is performed in the Global South with little control over its use \citep{gray2019ghost}, while African-led communities such as Masakhane have shown what participatory, locally controlled research can achieve \citep{nekoto2020participatory}. We show that the paradox of participation is reproduced specifically within evaluation infrastructure arrangements.

\subsection{Material Barriers to Evaluation Infrastructure}

Literature on compute access documents severe constraints in African contexts: only 5\% of Africa's AI talent has adequate computational resources \citep{undp2024compute}. \citet{solatorio2024double} document how speakers of low-resource languages incur higher financial and environmental costs per unit of useful output, compounded by currency volatility against provider pricing in US dollars, and most of the world's languages remain poorly served by the data and tools that evaluation depends on \citep{joshi2020state}. \citet{raji2021everything} document how benchmark availability shapes what can be evaluated. \citet{lehdonvirta2024compute} map a structural ``Compute North'' versus ``Compute South'' divide, in which countries without local AI compute face systematic disadvantages not only in AI development but in the regulatory and evaluative capacity that compute enables. Even systems described as open depend on infrastructure controlled by a small number of firms \citep{widder2024open}. Most of this work concerns infrastructure for development rather than evaluation and treats barriers one at a time; PAAI examines them together, as conditions of evaluation.

\subsection{Infrastructure as Sociotechnical Power}

Science and Technology Studies provides the theoretical foundation for understanding evaluation infrastructure as a power arrangement. \citet{star1999ethnography} argues that infrastructure is relational and political: invisible to those it serves well, and visible only when it breaks down or excludes. \citet{bowker1999sorting} demonstrate how standards and classification systems embody power relations, encoding particular assumptions about the world while marginalising others, and \citet{lampland2009standards} show how quantifying and formalising practices shape what can be known and by whom. Applied to AI evaluation, it shows that compute programmes, API terms and benchmark frameworks are not neutral conduits: they shape who can produce credible knowledge about AI systems and whose knowledge counts. \citet{crawford2021atlas} and \citet{sastry2024computing} extend this to the political economy of AI infrastructure, showing that control over compute functions as a governance lever, and \citet{luitse2024platform} traces how cloud platforms have become central to that political economy. \citet{whittaker2021steep} shows that researchers who depend on industry infrastructure face structural pressures that compromise the independence of their findings even without explicit censorship.

\section{The PAAI Framework}\label{sec:framework}

\subsection{The Gap This Framework Addresses}

Existing tools address adjacent but distinct questions. AEF-1 asks whether a specific evaluation met minimum conditions of access, resources, analytic autonomy and methodological transparency \citep{stosz2025aef1}; research on political economy characterises who controls AI development infrastructure \citep{crawford2021atlas,sastry2024computing}; and work on participation and structural capture explains how inclusion can coexist with unequal authority \citep{png2022tensions,whittaker2021steep}. AEF-1 presupposes an evaluator with sufficient access to be assessed. PAAI asks the prior question: do the infrastructure arrangements available to a given evaluator make adequate access achievable, on its own terms and over time? The answer determines whether capacity gaps are contingent, and so addressable through more resources, or structural, requiring changes to the terms and governance of infrastructure arrangements themselves.

\begin{figure*}[t]
\centering
\includegraphics[width=0.94\textwidth]{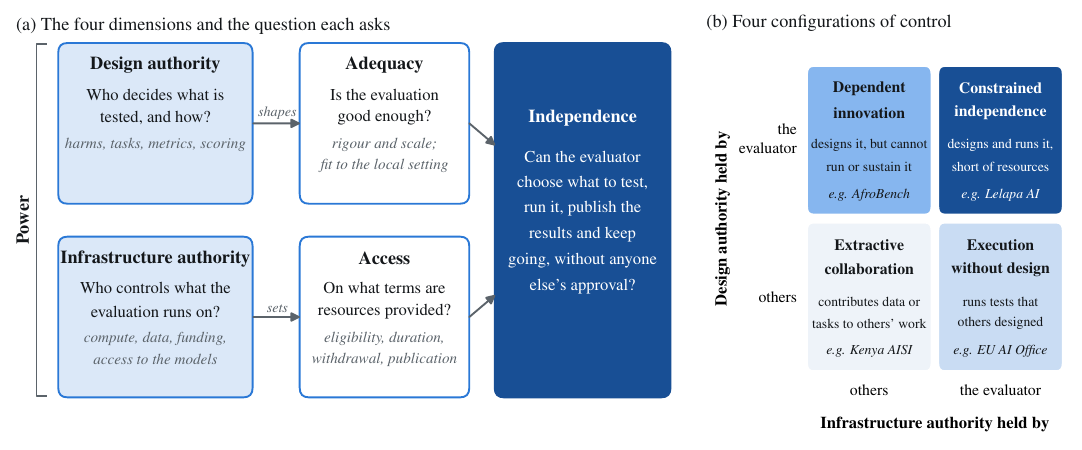}
\caption{The PAAI framework. (a) Power has two parts. Design authority decides what is tested and how, and shapes whether the evaluation fits its setting. Infrastructure authority controls the compute, data, funding and model access an evaluation runs on, and sets the terms of Access. Together they determine Independence. Each box gives the question PAAI asks. (b) Who holds each part of Power gives four configurations, each shown with an example from our data.}
\label{fig:framework}
\end{figure*}

\subsection{Theoretical Derivation of Each Dimension}

PAAI synthesises four scholarly traditions because evaluation infrastructure sits at their intersection and no single tradition covers it (Figure~\ref{fig:framework}a).

\paragraph{Power} draws on \citet{star1999ethnography} and \citet{bowker1999sorting}, whose work establishes that control over the material and institutional conditions enabling a practice translates into authority over what that practice can know and produce. We identify two sub-dimensions of Power. \emph{Design authority} concerns who defines evaluation frameworks, selects targets and tasks, sets priorities and authors findings. It operates at two levels: \emph{framework authority}, which sets the rules of an evaluation, such as what counts as harm, which metrics and automated scorers are used and how results are analysed; and \emph{content authority}, which contributes tasks, data or languages within a framework set by others. \emph{Infrastructure authority} concerns who controls the compute, data, funding and model access on which evaluation depends, and the terms under which they are provided. We code infrastructure authority as held by the evaluator where it controls the resources for the funded period and they cannot be withdrawn at a provider's discretion; philanthropic grants are therefore coded as evaluator-held and provider credits as partial. These sub-dimensions can be, and in African evaluation contexts frequently are, decoupled. \citet{casper2024blackbox} demonstrate that infrastructure authority shapes what evaluation is possible regardless of design authority: an evaluator who holds full intellectual control over evaluation design but depends on black-box API access is structurally prevented from executing rigorous evaluation of many safety-relevant properties.

\paragraph{Access} draws on the political economy of AI infrastructure \citep{crawford2021atlas,luitse2024platform}, which documents how access terms are not neutral technical arrangements but instruments through which providers shape what evaluation can occur. Access asks: under what conditions can actors use evaluation infrastructure? This includes eligibility criteria, geographic restrictions, time limits, renewal, usage constraints and publication requirements; whether the evaluator runs models itself or works on outputs produced by others; and whether access is fragmented across several short-term sources.

\paragraph{Adequacy} operationalises minimum standards from applied evaluation frameworks. AEF-1 holds that credible independent evaluation requires sufficient technical access (at minimum open-ended query access, ideally access to model internals), adequate compute for iterative and exploratory testing, sufficient time for thorough assessment, and analytic autonomy over evaluation scope, methods and the presentation of results \citep{stosz2025aef1}. We distinguish \emph{technical adequacy} (rigour and scale) from \emph{contextual validity} (whether an evaluation fits the deployment contexts it claims to inform) \citep{hutchinson2022evaluation}. Adequacy asks whether an evaluator's arrangements meet these conditions \citep{nist2023rmf}.

\paragraph{Independence} operationalises two complementary accounts of how infrastructure dependency compromises evaluator autonomy. Png's \citeyearpar{png2022tensions} paradox of participation identifies the dynamic whereby Global South actors are formally included in governance processes while remaining excluded from resource allocation and priority-setting. Whittaker's \citeyearpar{whittaker2021steep} account of structural capture shows how researchers who depend on industry infrastructure face implicit pressure to avoid findings that might jeopardise future access. Independence, as defined in the introduction, asks: can evaluators select evaluation targets without provider approval, publish findings without restriction, sustain evaluation programmes across model releases, and conduct adversarial assessment without risk to their infrastructure access? We assess Independence as this capacity and record recognition (co-authorship, credited sections) separately, because credit for contributing to an evaluation is not the same as the ability to conduct one.

\subsection{What the Four Dimensions Together Enable}

Applied alone, each dimension gives an incomplete picture: moderate Access with no design authority is a control problem, not a resource problem; African-led design with terminable access is a sustainability problem, not a design problem (Figure~\ref{fig:framework}a).

Crossing the two sub-dimensions of Power gives four configurations that look similar on the surface but have different root causes and require different remedies (Figure~\ref{fig:framework}b). Three emerge from the twelve African cases:
\begin{itemize}
\item \textbf{Extractive collaboration}: external or partial design authority with external infrastructure authority. African actors contribute labour, data or limited design inputs; Independence is minimal or absent.
\item \textbf{Dependent innovation}: African-led design authority with external or absent infrastructure authority. African actors lead evaluation design but cannot execute or sustain it without infrastructure they do not control.
\item \textbf{Constrained independence}: African-led design and African-led or partial infrastructure authority, limited by resources to scale or sustain evaluation.
\end{itemize}
A fourth configuration, \textbf{execution without design}, in which an actor runs tests that others designed, appears only in the joint testing data. Without all four dimensions, these configurations collapse into undifferentiated ``resource gaps'' with no clear remedy.

\subsection{Scope and Limitations of the Framework}

PAAI is designed specifically for evaluation infrastructure, not AI infrastructure generally. It applies whether an evaluator assesses its own models or others': in either case the question is whether the evaluator can choose, run, publish and continue its assessments without depending on those who supply its infrastructure. Independence from the developer of the assessed system matters additionally where evaluator and developer differ. Legitimacy, transparency and legal protections for auditors, identified by \citet{raji2022outsider} as prerequisites for consequential assessment, are not captured by PAAI and are directions for future extension. PAAI is an initial operationalisation, offered for refinement through engagement with practitioners conducting evaluation in resource-constrained settings.

\section{Methodology}\label{sec:methods}

\subsection{Project Identification}

We conducted structured documentary analysis of African AI evaluation projects carried out between November 2022 (the public release of ChatGPT) and December 2025. Sources searched included academic databases (ACM Digital Library, arXiv, ACL Anthology), organisational and policy reports, and public GitHub repositories. Search queries combined evaluation types with geographic and infrastructural terms (e.g.\ [evaluation type] AND ``Africa''; [country name] AND (``compute'' OR ``API'' OR ``benchmark'')). Of over 200 candidate sources, 47 were assessed for eligibility and 12 projects met all inclusion criteria.

Projects were included if they evaluated an identifiable object (model behaviour, benchmark performance or an evaluation framework), concerned African languages, institutions, users or deployment contexts, and were documented well enough to characterise at least two PAAI dimensions (Appendix~\ref{app:search}). Ten projects report empirical evaluation of models or benchmarks. Two, a critique of a medical benchmark \citep{mutisya2025healthbench} and a systematic review of AI in African higher education \citep{sangwa2025ethical}, are meta-evaluative: they assess evaluation frameworks or deployment evidence rather than running models. We retain them as boundary cases: they are among the few documented cases of African actors exercising full framework authority, and their independence is assessed for the work they undertook, which did not require running models. Excluding them leaves all three configurations populated. Lead institutions are in Kenya, Uganda, South Africa and Rwanda, alongside pan-African networks; four projects are led from Singapore and the UK, the United States, or Spain. The sample is a structured mapping of visible evaluation infrastructure arrangements, not a census.

\subsection{Joint Testing Exercises}

To test whether the patterns found in Africa reflect regional scarcity or the way shared evaluation is organised, we coded the role of every member institute in the three joint testing exercises published by the international network of AI Safety Institutes: a November 2024 pilot \citep{inais2024pilot}, a February 2025 multilingual and cybersecurity exercise \citep{inais2025paris,inais2025multilingual,vij2026multilingual}, and a July 2025 agentic exercise \citep{inais2025agentic,inais2025agenticreport,seah2026agentic}. The network was renamed the International Network for Advanced AI Measurement, Evaluation and Science in late 2025 \citep{aisi2026network}; we found no later published exercise. This yields 29 institute-by-exercise records (the United States did not take part in the third exercise). For each record we coded design authority, infrastructure authority (whether and how the institute ran models), access mode, adequacy issues for its language, and recognition; Appendix~\ref{app:jt} gives the categories and full coding.

\subsection{Thematic Analysis and Case Characterisation}

We applied thematic analysis \citep{braun2006thematic} using the PAAI dimensions as predefined themes, following a template-driven approach in which theoretically derived categories are applied systematically across cases. For each project, we extracted metadata, evaluation characteristics, infrastructure inputs, resource provision terms, scope and stated constraints. Where possible, information was triangulated across publications, project documentation, repositories and public statements. We also engaged informally with participants from several projects to check our reading of published documentation; these exchanges were not used as data. In parallel, we analysed the published eligibility, duration, renewal and publication terms of the access programmes the projects relied on (Appendix~\ref{app:providers}). Rather than assigning numerical scores, which would imply a precision inconsistent with qualitative evidence, we characterise each project with descriptive categories grounded in documentary evidence (Table~\ref{tab:cases}). Three researchers coded every case independently against a shared codebook and resolved disagreements through discussion until consensus, refining code definitions where needed \citep{mcdonald2019reliability}. Codebook, justifications and evidence for both datasets are given in Appendices~\ref{app:codebook}--\ref{app:jt}.

\begin{table*}[t]
\centering
\small
\begin{tabularx}{\textwidth}{@{}>{\raggedright\arraybackslash}p{3.3cm} l l >{\raggedright\arraybackslash}X >{\raggedright\arraybackslash}X >{\raggedright\arraybackslash}p{2.1cm} l@{}}
\toprule
\textbf{Project} & \textbf{Type} & \textbf{Power (D/I)} & \textbf{Access} & \textbf{Adequacy} & \textbf{Independence} & \textbf{Configuration} \\
\midrule
AISI Multilingual Joint Testing & Ml. safety & Ext / Ext & Mediated, time-limited & Externally defined & None & Extractive \\
Kenya AISI Agentic Testing & Ml. safety & Part / Ext & Coordinated; renewal not established & Limited relevance & Minimal & Extractive \\
Masakhane Uhura & Ml. safety & AL / Ext & Provider credits; renewal not established & Translated benchmarks & Partial & Dependent \\
Sunbird AI ASR & Benchmarking & AL / AL & Grant-funded & Contextually apt & High, cost-limited & Constrained \\
Lelapa InkubaLM & Benchmarking & AL / Part & Provider credits & Rigorous coverage & Partial & Constrained \\
Wits/UJ SAfriSenti & Benchmarking & AL / AL & Own hardware + cloud & Sufficient for scope & High, volunteer-limited & Constrained \\
Masakhane AfroBench & Benchmarking & AL / Ext & Multiple credits & Wide, English-adapted & Partial & Dependent \\
IrokoBench & Benchmarking & AL / Ext & Five short-term sources & Human translation & Partial & Dependent \\
Georgia Tech Medical LLM & Domain Risk & Ext / Ext & Data source only & Rigorous, external & Not assessed$^\ddagger$ & Extractive \\
Barcelona Fetal Ultrasound & Domain Risk & Ext / Ext & Data source only & Rigorous, external & Not assessed$^\ddagger$ & Extractive \\
Qhala HealthBench critique$^\dagger$ & Domain Risk & AL / None & Public documents & Proposal-stage & Of critique only & Dependent \\
Rwanda Higher-Ed review$^\dagger$ & Domain Risk & AL / None & Literature only & Policy-relevant & Of analysis only & Dependent \\
\bottomrule
\end{tabularx}
\caption{Twelve African AI evaluation projects characterised through PAAI. Power = design authority / infrastructure authority (AL = African-led; Ext = external; Part = partial). Configuration: extractive collaboration, dependent innovation or constrained independence. $^\dagger$Meta-evaluative case; configuration provisional, since no models were run. $^\ddagger$No African author in a design or analysis role. Justifications are in Appendix~\ref{app:cases}.}
\label{tab:cases}
\end{table*}

\section{Findings: Evaluation Infrastructure in Africa}\label{sec:africa}

\subsection{Multilingual Safety Testing}

Multilingual safety testing requires native-speaker expertise, culturally appropriate test scenarios and model access for behavioural assessment across languages \citep{bayes2024uhura,inais2025multilingual}. Three projects reveal a consistent pattern: African participation is high in linguistic labour and low in infrastructural control.

\paragraph{Power and control.} In the AISI multilingual exercise, Singapore led the exercise, ran all tests for ten languages on its own platform, developed the annotation guide and metrics calculator, and led the analysis. Harm categories were discussed among participating institutes, but the instruments that put them into practice were Singapore's; Kenya annotated and analysed Kiswahili \citep{inais2025multilingual}. Kenya's larger role in the agentic exercise is examined in Section~\ref{sec:jt}. Masakhane's Uhura benchmark was designed by an African research community but evaluated OpenAI and Anthropic models through commercial APIs, with free credits from OpenAI and several co-authors affiliated with a model provider \citep{bayes2024uhura}. This dependency confines evaluation to provider-exposed behaviour and excludes white-box analysis.

\paragraph{Access and adequacy.} Neither the Kenyan exercises nor Uhura documents renewable access that would support longitudinal safety monitoring: Kenya's access came through the exercise's coordination, with no documented right to renew it, and Uhura's API access rested on provider credits whose renewal terms are not established. Black-box access is already insufficient for rigorous evaluation of many safety-critical properties \citep{casper2024blackbox}; whether this limits a given evaluation depends on the properties it sets out to assess.

\paragraph{Independence and evaluation priorities.} Only one of the three projects introduced risks defined from African contexts: Kenya's fraud and phishing tasks in the agentic exercise, which entered an evaluation whose harm categories, scoring and analysis were set elsewhere. Kiswahili-specific failures in the multilingual exercise, such as models giving partial harmful instructions despite disclaimers, emerged as findings within an externally defined taxonomy rather than as part of a Kenyan agenda \citep{inais2025multilingual}. Uhura translated English benchmarks, risking the importation of Western evaluation assumptions \citep{hutchinson2022evaluation}. Without control over evaluation priorities, African deployment risks remain underrepresented in safety testing.

\subsection{Capability Benchmarking}

Capability benchmarking requires substantial compute, large-scale benchmark datasets and evaluation harnesses \citep{ojo2025afrobench,tonja2024inkubalm}. Five projects reveal different infrastructure positions within a common structural constraint.

\paragraph{Compute access.} Sunbird AI (Uganda) and Lelapa AI (South Africa) show the highest infrastructure capacity in the sample. Sunbird used H100/A100-class GPUs funded by a Hewlett Foundation grant, totalling about 20 GPU-hours for its data-scaling study \citep{akera2025asr}; Lelapa trained and benchmarked InkubaLM on eight A100 GPUs provided through free Microsoft Azure credits \citep{tonja2024inkubalm}. SAfriSenti used institutional hardware supplemented by Google Colab Pro+ and AWS EC2 instances, with no external funding \citep{mabokela2025safrisenti}, showing that constrained independence is achievable at modest cost, although reliance on volunteer annotators limits scale. AfroBench relied on two providers' credits and IrokoBench on five, each time-limited and separately applied for \citep{ojo2025afrobench,adelani2025irokobench}. None of the five documents compute access secured beyond the current project, which would be needed to keep benchmarking as new models are released.

\paragraph{Benchmark control.} Sunbird, Lelapa and the SAfriSenti team retained control over benchmark design, data and publication, the clearest examples of what African-led evaluation can achieve under present constraints. AfroBench achieved unprecedented coverage of 64 African languages but adapted many tasks from English datasets, raising questions about whether measured performance gaps reflect model failures or benchmark transfer artefacts \citep{peppin2025multilingual}; IrokoBench's human translation reduces but does not remove this concern.

\paragraph{Benchmarking after release.} None of the projects documents access to pre-deployment versions or to the internals of proprietary systems. The evidence they produce therefore arrives after release decisions have been made.

\subsection{Domain-Specific Risk Assessment}

Domain-specific risk assessment requires partnerships with domain institutions, contextually relevant data, and expertise bridging AI and specialised fields \citep{mutisya2025healthbench,sendra2023generalisability}. Four projects, two of them extractive collaborations, show both the starkest form of that pattern and its limits.

\paragraph{Data provision without evaluation capacity.} Hospitals in five African countries (Malawi, Egypt, Uganda, Ghana and Algeria) contributed fetal ultrasound scans to the Barcelona study, and ten clinicians and researchers from African institutions are credited co-authors for image collection \citep{sendra2023generalisability}. For the Georgia Tech and Google Research evaluation, knowledge of herbal remedies from ten African countries was drawn from published literature and WHO guidelines through desk research, and no authors were affiliated with African institutions \citep{nimo2025africa}. In neither case did African institutions design the evaluation or conduct the analysis. The Barcelona study ran on a single consumer-grade GPU, so the barrier was not hardware but the combination of expertise, institutional capacity and evaluation authority that the African sites were not positioned to supply. This is what \citet{couldry2019data} theorise as data colonialism, applied to AI knowledge production: knowledge leaves African contexts to inform evaluation conducted elsewhere, without a reciprocal transfer of evaluative capacity.

\paragraph{Qhala and Rwanda as partial exceptions.} Qhala's critique of HealthBench \citep{mutisya2025healthbench} and the Rwandan higher-education review \citep{sangwa2025ethical} show that African-led assessment of evaluation frameworks is possible without external infrastructure, but both remain at the level of critique and policy analysis rather than empirical evaluation. We classify both provisionally as dependent innovation: their autonomy is intellectual rather than operational, since the alternatives they propose cannot be implemented without infrastructure their authors do not control.

\begin{figure}[t]
\centering
\includegraphics[width=\columnwidth]{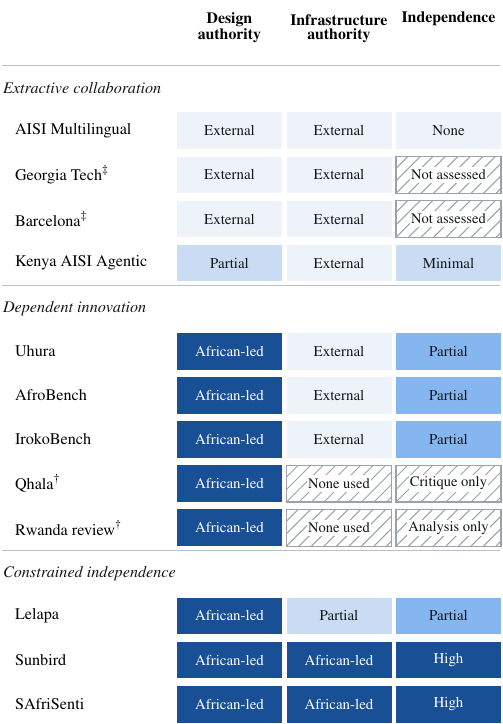}
\caption{Who held design and infrastructure authority in each African project, and the independence of African participants, grouped by configuration. Darker cells indicate more African authority or independence. Design authority alone never raised independence above partial. $^\dagger$ Meta-evaluative case (no models run); configuration provisional. $^\ddagger$ No African author in a design or analysis role (Georgia Tech: no African-affiliated authors; Barcelona: African co-authors credited for image collection only).}
\label{fig:configurations}
\end{figure}

\subsection{Cross-Cutting Patterns}

\paragraph{Independence follows infrastructure authority.} Figure~\ref{fig:configurations} sets out who held each kind of authority in every project. Where both were external, African participants in the AISI multilingual exercise had no documented independence; the Georgia Tech study identifies no African-affiliated author, and Barcelona's African co-authors are credited with image collection rather than design or analysis. Where Kenya shared in the design (the agentic exercise), independence was minimal. Where African actors held design but not infrastructure authority (Uhura, AfroBench, IrokoBench), or only partial infrastructure authority (Lelapa), independence was partial. Only Sunbird and SAfriSenti, which held both, reached high independence. Because the ability to run and repeat tests is part of what we mean by independence, its link to infrastructure authority is partly built into the definitions. What the cases add is that design authority alone never raised independence above partial. With this few cases, it is a documented pattern, not a statistical result.

\paragraph{Participation without power.} African actors contribute essential labour: task construction and language evaluation (Kenya), clinical data (Barcelona), domain knowledge through published literature (Georgia Tech), and annotation (SAfriSenti, AfroBench). Yet they frequently do not control evaluation frameworks, compute or research agendas. Contributing inputs is not equivalent to exercising authority over what is evaluated or how. This is Png's \citeyearpar{png2022tensions} paradox of participation reproduced within evaluation infrastructure.

\paragraph{Time-limited access as structural constraint.} Every project relying on external infrastructure faced temporal limits: API credits expire, cloud grants end and coordination windows close. This precarity prevents longitudinal monitoring and constrains adversarial testing, which requires independence from provider goodwill \citep{whittaker2021steep}.

\paragraph{Extractive knowledge flows and the AEF-1 floor.} Data and domain expertise flow from Africa to external evaluators, while evaluation capability remains concentrated elsewhere. Measured against AEF-1, the closest available standard \citep{stosz2025aef1}, none of the eight evaluations with an African evaluator documents all of its minimum operating conditions for African participants. Few academic evaluations anywhere would; what matters is which conditions fail and why. The closest, Sunbird and SAfriSenti, are limited by funding rather than by providers. The Georgia Tech and Barcelona studies, in which no African author held a design or analysis role, and the two meta-evaluative cases cannot be assessed against it. Extractive collaboration is incompatible with analytic autonomy, and time-limited access can suffice for one evaluation while leaving continuity unsecured.

\section{Findings: International Joint Testing}\label{sec:jt}

\begin{figure*}[t]
\centering
\includegraphics[width=0.94\textwidth]{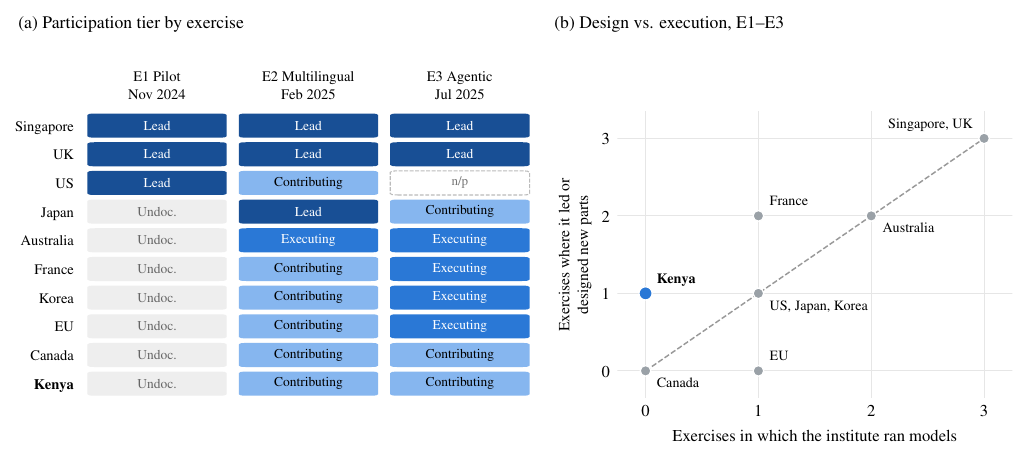}
\caption{Roles of the ten member institutes across the three published joint testing exercises. (a) Participation tier derived from coded design roles and documented model execution (n/p: not a participant; Undoc.: not documented, because the pilot report names only its leads). (b) Exercises with a lead or substantive design role against exercises in which the institute ran models; the dashed line marks equal counts.}
\label{fig:jointtesting}
\end{figure*}

Joint testing is currently the most visible route by which states share evaluation capacity, and its participants are mostly high-income. If PAAI captured only scarcity, well-resourced members would hold design and infrastructure authority together. Figure~\ref{fig:jointtesting} shows that several do not.

\paragraph{Leads run the models.} In every exercise, the lead institutes also executed the evaluation and built its analysis tools \citep{inais2025multilingual,inais2025agenticreport}. Framework authority and infrastructure authority coincide at the top of the arrangement.

\paragraph{Participants divide on execution.} Below the leads, institutes separate by whether they ran models. Australia ran independent tests in both later exercises; France, Korea and the EU AI Office ran tests or variations in the agentic exercise. Two members have no documented model execution in any exercise: Kenya, a lower-middle-income country, and Canada, a high-income country. Canada translated, annotated and wrote the analyses for Cantonese and Farsi and contributed to the cybersecurity analysis, a position close to Kenya's. The divide is therefore not reducible to national income. It reflects how joint testing allocates execution, although documents cannot tell us whether Canada's role was chosen or constrained.

\begin{table*}[t]
\centering
\small
\begin{tabularx}{\textwidth}{@{}>{\raggedright\arraybackslash}p{2.2cm} >{\raggedright\arraybackslash}p{3.6cm} >{\raggedright\arraybackslash}X >{\raggedright\arraybackslash}p{4.3cm}@{}}
\toprule
\textbf{Dimension} & \textbf{Diagnostic question} & \textbf{Design condition for joint testing and access programmes} & \textbf{Applied: Kenya, agentic exercise} \\
\midrule
Power: design authority & Who defines the harms, tasks and metrics, and how results are scored? & Each participant leads at least one harm category or task family, including its metrics and scoring. & Partly met: contributed new fraud and phishing tasks, but did not lead a strand or set scoring. \\
Power: infrastructure authority & Who controls the compute, data, funding and model access? & Each participant receives model access and compute directly, not only through the lead's platform. & Not met: tests were run on the lead's platform; no record of Kenya running tests itself. \\
Access & On what terms, and for how long? & Access terms, duration and renewal are written down, shared with participants and multi-year. & Not demonstrated in the public record: access was mediated through the exercise; duration and renewal terms are not reported. \\
Adequacy & Is the evaluation rigorous and fit for the contexts it informs? & Tasks reflect participants' deployment risks, and automated judges are validated per language. & Partly met: regional tasks included, but automated judges were least reliable for Kiswahili. \\
Independence & Can participants publish, replicate and continue the evaluation? & Participants may publish separately and keep using the datasets and tools they helped build. & Partly met: credited co-authorship; no documented right to rerun or reuse. \\
\bottomrule
\end{tabularx}
\caption{PAAI as a tool for diagnosis and design. The last column applies the questions to Kenya's role in the agentic joint testing exercise.}
\label{tab:design}
\end{table*}

\paragraph{Kenya: design authority grows, infrastructure authority does not.} In the multilingual exercise, Kenya annotated Kiswahili responses and authored the Kiswahili analysis. In the agentic exercise, it constructed new tasks and tools on fraud, phishing and impersonation, one of only two non-lead members to contribute new tasks, and also translated, annotated and contributed to the cybersecurity analysis \citep{inais2025agenticreport}. Kenyan staff are credited co-authors of both reports \citep{vij2026multilingual,seah2026agentic}. This expanded Kenya's content authority and brought regional risks into the evaluation. Yet the reports attribute the execution of Kenya's tasks to the lead, and scoring and analysis used the lead's instruments and framework. Of the four non-lead institutes that made substantive design contributions to the agentic exercise, Kenya is the only one that did not run models. The records show design authority without documented infrastructure authority, inside an arrangement intended to share capacity; they cannot show whether this division was chosen or constrained.

\paragraph{Agentic scoring was least reliable for participants' languages.} Automated judging in the agentic exercise was least reliable for Telugu, Hindi and Kiswahili \citep{inais2025agenticreport}. Telugu and Hindi were evaluated with no Indian member institution, and the pilot reported results for Swahili and Yoruba with no documented African role in that analysis \citep{inais2024pilot}. The instruments built by the institutes holding infrastructure authority performed worst on the languages of participants who held least of it, consistent with gaps documented in safety datasets for African languages \citep{onuoha2026gaps}. The records cannot show that the distribution of authority caused these failures, but they show that those best placed to detect them had least control over the instruments.

\section{Governance Propositions}\label{sec:props}

The configurations generate three governance propositions, each with an implication for how evaluation infrastructure is designed, funded and governed.

\subsection{Proposition 1: Access Without Control Can Reproduce Dependency}

The evaluations in which resources sat most fully with external institutions, the AISI joint testing exercises and the Barcelona and Georgia Tech studies, are those in which African actors held the weakest Power and Independence, because the institutions holding the resources also controlled framework design and priority-setting. Kenya's design role grew between exercises, yet the records give it no means to decide independently what counts as a safety concern in Kenyan contexts or to run that evaluation itself. This is a control gap that more resources, provided under the same terms, would not close.

Access that comes without authority over priorities, framework design, execution and publication reproduces participation without power, whatever its stated intentions. The condition for independence is not the volume of access but its governance: who decides what gets evaluated, by what methods, and who runs the evaluation.

\emph{Implication for joint testing coordinators and research partners:} participants should receive direct model access and lead at least one strand, with analytic autonomy and transparent publication rights as baseline conditions.

\subsection{Proposition 2: Project-Based Funding Is Poorly Suited to Independent Evaluation}

AEF-1 calls for sufficient time for thorough assessment, and credible oversight of deployed systems requires the institutional continuity to monitor them over time. Project-based funding can support high-quality individual evaluations, but it is poorly suited to sustaining the standing infrastructure that independent evaluation requires. All twelve African projects ran on time-limited resources: research grants, provider credits and coordination windows close when projects end. None documents standing evaluation infrastructure that combines dedicated compute, a permanent evaluation team and a multi-year mandate to assess AI systems deployed in African contexts; Kenya AISI has the latter two but, on the public record, not the first.

This is a structural mismatch, not simply a funding shortfall. Sunbird and Lelapa show that individual projects can achieve high-quality evaluation, but project funding does not by itself produce the capacities independent evaluation requires: monitoring models across updates, building expertise over time and accumulating knowledge of African deployment contexts. Unless something outlasts the project, capability ends with it. For frontier AI developers and cloud providers, the corresponding condition is multi-year, renewable access, including access to model internals where an evaluation requires it.

\emph{Implication for international bodies and governments:} the first wave of AI Safety Institutes concentrated standing evaluation capacity in a small number of high-income countries \citep{araujo2024aisis}. International bodies should support institutional, not merely project-based, evaluation infrastructure in Africa: multi-year funding, dedicated compute and authority over evaluation priorities.

\subsection{Proposition 3: Evaluation Framework Co-Design Is a Governance Problem, Not Only a Methodological One}

When evaluation frameworks are defined externally, by AI Safety Institutes in other regions, by Western academic institutions or by AI providers, African evaluators cannot determine what counts as harm, risk or adequate performance in their deployment contexts. Uhura translated English benchmarks rather than developing benchmarks from African deployment realities; Kenya's own tasks were scored with instruments designed elsewhere; AfroBench adapted tasks from English datasets. These are not only translation problems; they reflect a governance arrangement in which authority over evaluation design sits outside the contexts where AI systems are deployed.

Models produce more harmful content in low-resource languages \citep{yong2024lowresource,peppin2025multilingual}, yet standard safety evaluations rarely test for these disparities \citep{deng2024multilingual}, and in the agentic joint testing exercise the automated scorers were least reliable on exactly these languages. Such failures are predictable when frameworks do not include African practitioners in priority-setting. Framework co-design is therefore a prerequisite for the legitimacy of evaluation evidence in AI governance, not a procedural nicety.

\emph{Implication for standard-setters and evaluation designers:} co-design with resource-constrained evaluators should be treated as a central condition of AI accountability. Minimum standards for independent evaluation, including AEF-1, have so far been drafted primarily by organisations that already have the infrastructure to meet them, and risk encoding assumptions that do not hold in resource-constrained settings.

 Table~\ref{tab:design} turns each PAAI question into a design condition that can be checked against the published record of any exercise or programme, and applies it to Kenya in the agentic exercise.

\section{Discussion}

\subsection{Situating the Findings}

The patterns identified here reproduce, in evaluation, dynamics documented across the AI pipeline \citep{birhane2020algorithmic,couldry2019data,kwet2019digital,png2022tensions}. PAAI specifies how they operate in evaluation infrastructure. The same structure appears in joint testing among mostly high-income members, which supports reading independence as a property of infrastructure arrangements rather than of evaluators alone. The documented gaps concern the conditions for sustained independent scrutiny, not only equity.

\subsection{Limitations}

First, both datasets are documentary: the twelve projects are those identifiable from public documentation that met our criteria (Appendix~\ref{app:search}), not a representative sample, and because reports may describe leads' roles more fully than participants', ``not documented'' means unreported, not absent. Second, documents cannot distinguish chosen from constrained non-execution: Canada, and possibly Kenya, may have specialised by agreement. Third, proprietary contract terms are undisclosed, so Access and Independence rest on observable indicators. Fourth, AEF-1 was designed for third-party evaluations of developers' systems and fits some of our cases imperfectly. Fifth, PAAI characterisations are qualitative judgements, reached by consensus among three coders (Section~\ref{sec:methods}) rather than through inter-rater statistics; the evidence for each is in Appendices~\ref{app:cases} and~\ref{app:jt}. Sixth, two of the three joint testing exercises also appear among the African cases through Kenya's participation, so the comparison extends the African analysis rather than replicating it independently.

\subsection{Future Work}

We plan to extend this study through fieldwork and interviews with practitioners in Africa and with joint testing participants, to establish how roles are negotiated and to refine PAAI with those who would use it. We also plan to apply PAAI to third-party evaluators in high-income countries and in other resource-constrained regions.

\section{Conclusion}

This paper asked not whether African actors participate in AI evaluation but whether the infrastructure arrangements they depend on provide the forms of control required for independent evaluation. Across twelve African projects in multilingual safety testing, capability benchmarking and domain-specific risk assessment, and across three international joint testing exercises, the answer is consistent: existing arrangements expand participation more readily than they transfer evaluative authority.

Independent evaluation capacity requires institutional infrastructure rather than access programmes, co-designed frameworks, and sustained, evaluator-controlled compute and model access rather than time-limited credits.

\section*{Ethical Statement}

This study relies on public documentation. Informal exchanges with practitioners were used only to check our reading; no personal data were collected and no interview data are reported. Our characterisations concern infrastructure arrangements, not the quality of anyone's work. The main risk of adverse impact is that the critique is read as criticising collaboration or particular institutes; we argue that collaboration is necessary and should build independent capacity, not substitute for it.

\bibliography{references}

\appendix
\section*{Appendices: Data and Coding}

These appendices document the two datasets used in the paper. Appendix~\ref{app:codebook} explains what we looked for under each PAAI question. Appendix~\ref{app:search} explains how we chose the cases. Appendix~\ref{app:cases} gives the evidence behind each entry for the twelve African projects. Appendix~\ref{app:providers} summarises the access programmes those projects relied on. Appendix~\ref{app:jt} records each institute's role in the joint testing exercises. The full dataset, a spreadsheet with six sheets, will be released publicly if the paper is accepted. These appendices contain the same evidence and rules, so the paper can be assessed without it. All sources are public documents, and no personal data are included.

\section{What We Looked For}\label{app:codebook}

\paragraph{Design authority.} We asked who defined the harms, measures and scoring; who wrote the annotation guides; who chose what to test; and who led the analysis and wrote up the findings. \emph{Framework authority} means setting these rules. \emph{Content authority} means contributing tasks, data or languages within rules set by others. We drew on author affiliations, method and contribution sections, and acknowledgements.

\paragraph{Infrastructure authority.} We asked who controlled the computing power and model access; whether the team used models directly or through someone else; and whether a provider or lead institution could withdraw access. We drew on acknowledgements of computing support, provider terms, and descriptions of where and by whom models were run.

\paragraph{Access.} We asked whether the team ran models itself or worked with results produced by others; who was eligible and where; how long access lasted and whether it could be renewed; how many separate sources the team relied on; and whether it depended on a go-between. We drew on provider programme pages and acknowledgements.

\paragraph{Adequacy.} For \emph{technical adequacy}, we asked whether the computing power, data and model access matched what the evaluation claimed, and whether the team could rerun and refine its tests. For \emph{contextual validity}, we asked whether the tasks, languages and measures reflected the settings the evaluation claimed to describe, whether tasks were built locally or translated, and how reliable automated scoring was in each language. We drew on descriptions of scope, benchmark documentation, stated limitations and reported results.

\paragraph{Independence.} We asked whether the team could choose what to evaluate, design the methods, repeat the evaluation, publish and keep going over time without needing approval or goodwill from providers or lead institutions. Being credited as an author or having a named section was recorded separately and did not by itself raise the level. For the ten African projects that ran evaluations, we used four levels: none (no documented control over targets, execution or publication), minimal (contributes content but cannot run or continue the evaluation), partial (controls design and publication but depends on infrastructure that can be withdrawn) and high (controls design, execution and publication, limited only by resources). We drew on publication practices, funding arrangements, statements about provider dependence, and evidence that work was repeated or continued.

\paragraph{Configurations.} We placed each case in one of four configurations according to who held the two sub-dimensions of Power. A team holding both design and infrastructure authority is in \emph{constrained independence}. A team holding design but not infrastructure authority is in \emph{dependent innovation}. A team that runs tests others designed is in \emph{execution without design}, which appears only in joint testing. A team holding neither is in \emph{extractive collaboration}.

\section{How We Chose the Cases}\label{app:search}

\paragraph{Search.} Academic databases (ACM Digital Library, arXiv, ACL Anthology), organisational and policy reports, and public repositories, November 2022 to December 2025. Queries combined evaluation types with geographic and infrastructural terms, e.g.\ [evaluation type] AND ``Africa''; [country name] AND (``compute'' OR ``API'' OR ``benchmark''). The searches returned more than 200 sources. After removing duplicates and screening, we assessed 47 in detail and included 12 projects.

\paragraph{Inclusion.} A project was included if (1) it evaluated something identifiable, such as a model's behaviour, performance on a benchmark, or an evaluation method itself; (2) it concerned African languages, institutions, users or settings; (3) its public documents were detailed enough to answer at least two PAAI questions; and (4) it took place within the date range.

\paragraph{Exclusion.} We excluded (1) evaluations of Western settings with no African focus, (2) proposals that did not evaluate anything identifiable, and (3) studies of under-resourced languages that were not African.

\paragraph{Special cases.} Two projects (C.9 and C.10) assessed evaluations or published evidence without running models. InkubaLM (C.5) mainly built a model and is included because it also tested that model on six tasks. Removing C.9 and C.10 still leaves cases in all three African configurations.

\paragraph{Joint testing.} We included all three joint testing exercises the network had published by October 2026, and recorded the role of every member institute named in each report. The pilot report names only its three leads.

\section{Evidence for Each African Project}\label{app:cases}

Each subsection matches a row of Table~\ref{tab:cases}. The line under each heading gives the type of evaluation and the configuration. The ``AEF-1'' notes compare each case with the closest available standard \citep{stosz2025aef1}, which was written for outside evaluations of companies' systems and fits several cases imperfectly.

\subsection*{C.1 AISI Multilingual Joint Testing}
\emph{Multilingual safety $\cdot$ Extractive collaboration}

\textbf{Power (External / External).} Singapore AISI led the exercise, tested two open-weight models (Mistral Large and Gemma 2 27B) on its Moonshot platform for all ten languages, developed the annotation guide and automated metrics calculator, and led the analysis. Harm categories were discussed among participating institutes, but Singapore built the tools that applied them. Kenya annotated Kiswahili responses and authored the Kiswahili analysis section. \textbf{Access (Through the lead, time-limited).} Kenya worked on results produced on Singapore's platform during the exercise; no direct model access is recorded. \textbf{Adequacy (Rules set externally).} Rigorous methods: over 6,000 translated prompts, scored by an AI model and checked by human annotators. Kiswahili showed up to 31 percentage points of variation between the two tested models, and models sometimes gave partial harmful instructions despite disclaimers; these emerged as findings within a list of harms defined by others. \textbf{Independence (None documented).} Kenya did not choose the models, define the measures or repeat the evaluation, and the records show no right to do so. \textbf{Recognition:} named Kiswahili section; Kenyan staff are credited co-authors of the arXiv version. \textbf{AEF-1:} freedom of analysis and independent access not met. \textbf{Sources:} \citet{inais2025multilingual,inais2025paris,vij2026multilingual}.

\subsection*{C.2 Kenya AISI Agentic Testing}
\emph{Multilingual safety $\cdot$ Extractive collaboration}

\textbf{Power (Partial / External).} Kenya constructed a new set of ten tasks and tools on fraud, phishing and impersonation, making it one of two non-lead institutes to contribute new tasks. It also translated the dataset into Kiswahili, annotated results and contributed to analysing the cybersecurity test records. Singapore led the data leakage and fraud part, ran the tests on Moonshot, supplied the annotation guides and led the analysis; the UK led the cybersecurity part. \textbf{Access (Coordinated; renewal not established).} The lead ran Kenya's tasks on its own platform; no Kenyan model runs or computing resources are recorded. \textbf{Adequacy (Limited fit to context).} Kenya's tasks brought in regional risks, but scoring and analysis followed the lead's framework, and the automated scorers disagreed most with human reviewers on Telugu, Hindi and Kiswahili. \textbf{Independence (Minimal).} Kenya contributed substantive tasks and analysis, while the lead institutes retained responsibility for the overarching framework and execution. The records do not establish Kenya’s authority to initiate further runs or continue the evaluation independently. \textbf{Recognition:} Kenyan staff are credited co-authors of the arXiv version. \textbf{AEF-1:} independent access not met; the rules were set by others. \textbf{Sources:} \citet{inais2025agentic,inais2025agenticreport,seah2026agentic}.

\subsection*{C.3 Georgia Tech / Google Research Medical LLM Audit}
\emph{Domain-specific risk $\cdot$ Extractive collaboration}

\textbf{Power (External / External).} Georgia Tech and Google Research designed the Cultural Bias Score and Cultural Bias Attribution metrics, ran all the analysis, including analysis of the models' internal workings, and wrote the paper. Knowledge of herbal remedies from ten African countries came from PubMed (2020--2024) and WHO guidelines. \textbf{Access (Data source only).} No African institution used the computing resources, models or tools. \textbf{Adequacy (Rigorous, external).} New measures and detailed model analysis on computing resources the paper does not describe. \textbf{Independence (not assessed).} No authors affiliated with African institutions are identified, so the case informs who held evaluative authority rather than the independence of an African evaluator. \textbf{AEF-1:} not applicable. \textbf{Sources:} \citet{nimo2025africa}.

\subsection*{C.4 Sunbird AI ASR Data Scaling}
\emph{Capability benchmarking $\cdot$ Constrained independence}

\textbf{Power (African-led / African-led).} Sunbird AI (Uganda) designed the study, curated data from a 1,400-hour Kinyarwanda corpus and Kikuyu data, trained and evaluated models, published and released models openly. The funder provided resources without design authority. \textbf{Access (Grant-funded).} High-end graphics processors paid for by a Hewlett Foundation grant, used for about 20 hours in total. \textbf{Adequacy (Good fit to context).} 270 hours of curated training audio and 6,910 test samples, with analysis of errors under real conditions. \textbf{Independence (High, cost-limited).} No provider limits on publication; continuing depends on grant renewal rather than on a provider. \textbf{AEF-1:} closest to the standard; constrained by funding. \textbf{Sources:} \citet{akera2025asr}.

\subsection*{C.5 Lelapa AI InkubaLM}
\emph{Capability benchmarking $\cdot$ Constrained independence}

\textbf{Power (African-led / Partial).} Lelapa AI (South Africa) defined the problem, curated Inkuba-Mono (2.4B tokens, five languages) and instruction data, trained the 0.4B-parameter model, tested it on six tasks and kept ownership of the model; international partners provided research support. \textbf{Access (Provider credits).} Eight high-end graphics processors through free Microsoft Azure credits, which Microsoft could withdraw; venture funding offers some protection. \textbf{Adequacy (Rigorous coverage).} Performs comparably to larger models on the reported tasks. \textbf{Independence (Partial).} Published independently and kept ownership, but depends on provider computing. \textbf{AEF-1:} partly limited by dependence on provider computing. \textbf{Sources:} \citet{tonja2024inkubalm}.

\subsection*{C.6 Barcelona Fetal Ultrasound Generalisation}
\emph{Domain-specific risk $\cdot$ Extractive collaboration}

\textbf{Power (External / External).} The University of Barcelona and BCNatal trained an image model on Spanish data, checked it on Danish data, developed the method for adapting it to new hospitals and tested it on African data; all computation and analysis took place in Spain. \textbf{Access (Data source only).} Hospitals in Egypt, Uganda, Ghana, Malawi and Algeria collected ultrasound data (120 patients in total) and did not take part in model design, training or analysis. \textbf{Adequacy (Rigorous, external).} Trained on a single standard graphics card; after adaptation using a few local patients, the model was highly accurate on African data (AUC above 0.98). \textbf{Independence (not assessed).} Development and analysis were located in Spain. Ten co-authors from African institutions are credited with image collection; the paper documents no design, training or analysis role at the African sites. Recognition: co-authorship without evaluative authority. \textbf{AEF-1:} not applicable to African participants. \textbf{Sources:} \citet{sendra2023generalisability}.

\subsection*{C.7 Wits/UJ SAfriSenti}
\emph{Capability benchmarking $\cdot$ Constrained independence}

\textbf{Power (African-led / African-led).} South African universities designed the study, curated over 115,000 tweets in five Southern African languages, designed a partly automated labelling process with native-speaker volunteers, adapted existing multilingual models and released the collection. \textbf{Access (Own hardware + cloud).} University computers plus paid cloud services (Google Colab Pro+ and AWS); no specific funding. \textbf{Adequacy (Sufficient for scope).} Labelling method can be repeated; scale limited by reliance on volunteers. \textbf{Independence (High, volunteer-limited).} No funder restrictions; collection publicly released. \textbf{AEF-1:} closest to the standard; limited by resources rather than by providers. \textbf{Sources:} \citet{mabokela2025safrisenti}.

\subsection*{C.8 Masakhane Uhura}
\emph{Multilingual safety $\cdot$ Dependent innovation}

\textbf{Power (African-led / External).} The Masakhane community designed the benchmark and human-translated ARC-Easy and TruthfulQA into six African languages. Models were tested through the companies' online services: GPT-4o and o1-preview (OpenAI) and Claude (Anthropic), with free credits from OpenAI; several co-authors worked for OpenAI. This is the clearest case in the sample of a team controlling the design but not the infrastructure. \textbf{Access (Free online access, can be withdrawn).} The team saw only the models' inputs and outputs, and the providers could end access. \textbf{Adequacy (Translated benchmarks).} Rigorous, but built on English originals. \textbf{Independence (Partial).} Published independently, but seeing only inputs and outputs limits how hard the team could probe the models, and losing access would stop the work. \textbf{AEF-1:} sustained access not established. \textbf{Sources:} \citet{bayes2024uhura}.

\subsection*{C.9 Qhala HealthBench Critique$^\dagger$}
\emph{Domain-specific risk $\cdot$ Dependent innovation (provisional)}

\textbf{Power (African-led / None).} Kenyan authors identified flaws in OpenAI's HealthBench (reliance on expert opinion over stronger forms of evidence, neglect of tropical diseases, and Western clinical assumptions) and proposed an alternative based on clinical guidelines. \textbf{Access (Public documents).} Published HealthBench documentation and Kenyan, South African and WHO guidelines; no model or compute access. \textbf{Adequacy (Proposal only).} No evaluation was run. \textbf{Independence (of critique only).} Published without funder constraints; the documents do not show whether the authors had or sought resources to implement the alternative. \textbf{AEF-1:} not assessable. \textbf{Sources:} \citet{mutisya2025healthbench}.

\subsection*{C.10 Rwanda Higher-Education Review$^\dagger$}
\emph{Domain-specific risk $\cdot$ Dependent innovation (provisional)}

\textbf{Power (African-led / None).} Led from the African Leadership University (Rwanda) with a US co-author institution; a systematic review of published studies on AI in African universities, ending in a plan for how universities should govern AI. \textbf{Access (Literature only).} \textbf{Adequacy (Policy-relevant).} Quantified effects are those reported in the reviewed studies; no technical evaluation. \textbf{Independence (of analysis only).} The authors controlled their analysis; implementing the plan would need resources the documents do not describe. \textbf{AEF-1:} not assessable. \textbf{Sources:} \citet{sangwa2025ethical}.

\subsection*{C.11 Masakhane AfroBench}
\emph{Capability benchmarking $\cdot$ Dependent innovation}

\textbf{Power (African-led / External).} Masakhane aggregated 22 datasets across 64 African languages and 15 tasks and designed how they would be tested; international partners provided infrastructure for running the tests. Computing came from Google Cloud and OpenAI credits. \textbf{Access (Multiple credits).} Time-limited credits; a smaller AfroBench-Lite was created so that others could run it more cheaply. \textbf{Adequacy (Wide, English-adapted).} Wider coverage than any earlier benchmark for African languages; many tasks adapted from English sources. \textbf{Independence (Partial).} Published independently with a public leaderboard; continuing depends on credits. \textbf{AEF-1:} sustained access not established. \textbf{Sources:} \citet{ojo2025afrobench}.

\subsection*{C.12 IrokoBench}
\emph{Capability benchmarking $\cdot$ Dependent innovation}

\textbf{Power (African-led / External).} Masakhane-led human translation of three tasks (AfriXNLI, AfriMGSM, AfriMMLU) into 17 African languages, with many African partner institutions. \textbf{Access (Five short-term sources).} Lacuna Fund, Oracle compute credits, OpenAI Researcher Access Program, Google's Gemma academic programme and a Cohere For AI grant; the most scattered set of sources in the sample. \textbf{Adequacy (Human translation).} Highest translation quality among the benchmarking cases. \textbf{Independence (Partial).} Published independently; no long-term computing arrangement. \textbf{AEF-1:} sustained access not established. \textbf{Sources:} \citet{adelani2025irokobench}.

\section{Provider Access Programmes}\label{app:providers}

Table~\ref{tab:providers} summarises the published terms of the programmes the African projects used. Private contract terms were not available. None of the programmes offers long-term access that renews automatically. All are tied to a project or a fixed amount of credit, and researchers must apply again. None of the commercial programmes publishes its rules on what researchers may publish.

\begin{table*}[t]
\centering\footnotesize

\begin{tabularx}{\textwidth}{@{}>{\raggedright\arraybackslash}p{2.2cm} >{\raggedright\arraybackslash}p{3.4cm} >{\raggedright\arraybackslash}X >{\raggedright\arraybackslash}p{3.2cm} >{\raggedright\arraybackslash}p{2.4cm}@{}}
\toprule
\textbf{Provider} & \textbf{Programme} & \textbf{Duration and renewal} & \textbf{Publication terms} & \textbf{Projects} \\
\midrule
OpenAI & Researcher Access Program & Project-based; duration not disclosed; reapplication at provider discretion & Not disclosed; usage policy applies & C.8, C.11, C.12 \\
Anthropic & External Researcher Access Program & Project-based; reapplication required & Not disclosed; usage policy applies & C.8 \\
Google & Cloud research credits; Gemma academic programme; Lacuna Fund co-funder & Credits expire per cycle; competitive reapplication & Not disclosed & C.11, C.12 \\
Microsoft & Azure credits (direct provision) & Duration of credit agreement; reapplication required & Not disclosed & C.5 \\
AWS & Research credits & 12-month cycles; competitive reapplication & Not disclosed & None (C.7 used paid EC2) \\
Oracle & Oracle for Research credits & Project-based; application-based & Not disclosed & C.12 \\
Cohere For AI & Research grant & Grant cycle; competitive & Not disclosed & C.12 \\
Lacuna Fund & Dataset grants (Rockefeller Foundation, Google.org, IDRC) & Grant cycle; competitive & Open data and public results required & C.12 \\
Hewlett Foundation & Programme grant & Grant cycle; multi-year possible & Generally permissive for academic publication & C.4 \\
Cassava Technologies & No documented AI research access programme & -- & -- & None \\
\bottomrule
\end{tabularx}
\caption{Infrastructure access programmes used by the African projects (public documentation, accessed December 2025). Numbers refer to Appendix~\ref{app:cases}.}
\label{tab:providers}
\end{table*}

\section{Joint Testing Records}\label{app:jt}

\paragraph{What each record contains.} There is one record for each member institute in each exercise (29 records), combining the separate parts of an exercise. \emph{Design} is one of: Lead (led the exercise or one of its parts, or wrote its core method, annotation guides or measures); New parts (contributed new tasks, tools, benchmarks, measures, scoring prompts or its own test variations); Language/analysis (translation, checking, annotation or analysis only); None recorded. \emph{Running models} is one of: Main tests (ran the main tests); Own tests (ran its own tests, repeats or variations); None recorded. \emph{Access} is Direct (ran models) or Via lead (worked with results produced by another institute). \emph{Role} follows from these: Lead if the institute led; otherwise Executing if it ran models; otherwise Contributing if any role is recorded; otherwise Undocumented. \emph{Evidence} is High if the report states the role, Medium if it comes from a summary of the report, and Low if the institute is not named.

\paragraph{First exercise (pilot, November 2024).} The report names three leads, the United States, United Kingdom and Singapore, which tested Llama-3.1 405B on three benchmarks; the United States ran the 14-language MMMLU evaluation, which found English substantially outperforming Yoruba and Swahili. Other members contributed ``input'' but are not named, so their roles are recorded as undocumented \citep{inais2024pilot}.

\paragraph{Second and third exercises.} Tables~\ref{tab:jte2} and~\ref{tab:jte3} give the records. Sources: \citet{inais2025paris,inais2025multilingual,vij2026multilingual} for the second exercise; \citet{inais2025agentic,inais2025agenticreport,seah2026agentic} for the third.

\begin{table}[!ht]
\centering\footnotesize

\begin{tabular}{@{}l c c c c@{}}
\toprule
\textbf{Institute} & \textbf{Roles} & \textbf{Led} & \textbf{Ran models} & \textbf{Design} \\
\midrule
Singapore & 3 & 3 & 3 & 3 \\
United Kingdom & 3 & 3 & 3 & 3 \\
United States & 2 & 1 & 1 & 1 \\
Japan & 2 & 1 & 1 & 1 \\
Australia & 2 & 0 & 2 & 2 \\
France & 2 & 0 & 1 & 2 \\
Korea & 2 & 0 & 1 & 1 \\
EU AI Office & 2 & 0 & 1 & 0 \\
Canada & 2 & 0 & 0 & 0 \\
Kenya & 2 & 0 & 0 & 1 \\
\bottomrule
\end{tabular}
\par\smallskip\raggedright
Roles: exercises with a recorded role. Design: exercises in which the institute led or designed new parts.
\caption{Institute totals across the three exercises.}
\label{tab:jtsummary}
\end{table}

\paragraph{Totals and scoring quality.} Table~\ref{tab:jtsummary} gives each institute's totals, which are the values plotted in Figure~\ref{fig:jointtesting}b. In the second exercise, the automated scorer disagreed with human reviewers more than 5\% of the time in six of ten languages, most often for Japanese (18.9\%) and Telugu (15.0\%) \citep{inais2025multilingual}. In the third, disagreement between automated scorers and humans ranged from 15\% to 41\%, depending on the scoring model, and was highest for Telugu, Hindi and Kiswahili in most risk categories \citep{inais2025agenticreport}.

\paragraph{Limits of these records.} Reports may describe the leads' work more fully than the participants'. ``None recorded'' means the report does not mention a role, not that there was none. We took roles from the published reports and summaries. Author counts are estimated from the order of names on the arXiv versions, which list the lead institutes first and then other members alphabetically by country.

\begin{table*}[p]
\centering\footnotesize\setlength{\tabcolsep}{3pt}
\begin{tabularx}{\textwidth}{@{}>{\raggedright\arraybackslash}p{1.6cm} >{\raggedright\arraybackslash}X >{\raggedright\arraybackslash}p{1.7cm} >{\raggedright\arraybackslash}p{1.6cm} >{\raggedright\arraybackslash}p{1.5cm} >{\raggedright\arraybackslash}p{1.65cm} >{\raggedright\arraybackslash}p{2.4cm} >{\raggedright\arraybackslash}p{1.35cm}@{}}
\toprule \textbf{Institute} & \textbf{Documented role} & \textbf{Design} & \textbf{Ran models} & \textbf{Access} & \textbf{Role} & \textbf{Credit} & \textbf{Evidence} \\ \midrule
Singapore & Exercise lead (multilingual); ran all tests; wrote annotation guide and metrics calculator; led analysis & Lead & Main tests & Direct & Lead & 12 authors; First-listed author block & High \\
Japan & Co-lead (multilingual); translated dataset into ten languages; ran independent tests; Japanese annotation and language analysis; translated cyber benchmarks & Lead & Own tests & Direct & Lead & 5 authors; Japanese language analysis & High \\
UK & Lead of cybersecurity strand; ran cyber evaluations & Lead & Main tests & Direct & Lead & 4 authors & High \\
Australia & Ran independent tests in Mandarin Chinese; input on cyber analysis & New parts & Own tests & Direct & Executing & 7 authors & High \\
France & Contributed internal cyber benchmark; French annotation and language analysis & New parts & None recorded & Via lead & Contributing & 3 authors; French language analysis & High \\
Korea & Korean annotation and language analysis; Korean translation of cyber benchmarks & Language/ analysis & None recorded & Via lead & Contributing & 7 authors; Korean language analysis & High \\
Canada & Translated/validated Cantonese and Farsi; annotation; two language language analyses & Language/ analysis & None recorded & Via lead & Contributing & 4 authors; Cantonese and Farsi language analyses & Medium \\
Kenya & Kiswahili annotation and language analysis & Language/ analysis & None recorded & Via lead & Contributing & 4 authors; Kiswahili language analysis & High \\
EU AI Office & Listed as joint participant; input on discussions and cyber analysis & Language/ analysis & None recorded & Not recorded & Contributing & None & Low \\
US & Input on cyber analysis & Language/ analysis & None recorded & Not recorded & Contributing & None & Low \\
\bottomrule\end{tabularx}
\caption{Second exercise (multilingual and cybersecurity, February 2025). Column meanings are given in Appendix~\ref{app:jt}. Author counts are estimated from the order of names on the arXiv version.}
\label{tab:jte2}
\end{table*}
\begin{table*}[p]
\centering\footnotesize\setlength{\tabcolsep}{3pt}
\begin{tabularx}{\textwidth}{@{}>{\raggedright\arraybackslash}p{1.6cm} >{\raggedright\arraybackslash}X >{\raggedright\arraybackslash}p{1.7cm} >{\raggedright\arraybackslash}p{1.6cm} >{\raggedright\arraybackslash}p{1.5cm} >{\raggedright\arraybackslash}p{1.65cm} >{\raggedright\arraybackslash}p{2.4cm} >{\raggedright\arraybackslash}p{1.35cm}@{}}
\toprule \textbf{Institute} & \textbf{Documented role} & \textbf{Design} & \textbf{Ran models} & \textbf{Access} & \textbf{Role} & \textbf{Credit} & \textbf{Evidence} \\ \midrule
Singapore & Lead of leakage/fraud strand; ran tests; annotation guides; led analysis & Lead & Main tests & Direct & Lead & 12 authors; First-listed author block & High \\
UK & Lead of cybersecurity strand; ran evaluations; statistical framework & Lead & Main tests & Direct & Lead & 11 authors; Second-listed author block & High \\
France & New tasks and tools; ran variant tests; alternate judge prompt; extra metrics; French translation and annotation & New parts & Own tests & Direct & Executing & 6 authors; French analysis & High \\
Korea & Validated annotations; ran temperature/model-size variations; translation, annotation; analysis of cyber test records & New parts & Own tests & Direct & Executing & 7 authors; Korean analysis & High \\
Australia & Validated annotations; added metrics; ran cyber evaluations; translation, annotation; analysis of test records & New parts & Own tests & Direct & Executing & 14 authors & High \\
EU AI Office & Ran cyber evaluations; analysis of test records & Language/ analysis & Own tests & Direct & Executing & 4 authors & High \\
Japan & Validated annotations; Japanese translation, annotation and language analysis & Language/ analysis & None recorded & Via lead & Contributing & 3 authors; Japanese analysis & High \\
Canada & Farsi translation, annotation and language analysis; analysis of cyber test records & Language/ analysis & None recorded & Via lead & Contributing & 9 authors; Farsi analysis & High \\
Kenya & New fraud/phishing tasks and tools (10 tasks); Kiswahili translation, annotation and language analysis; analysis of cyber test records & New parts & None recorded & Via lead & Contributing & 4 authors; Kiswahili analysis & High \\
\bottomrule\end{tabularx}
\caption{Third exercise (agentic, July 2025). Column meanings are given in Appendix~\ref{app:jt}. Author counts are estimated from the order of names on the arXiv version.}
\label{tab:jte3}
\end{table*}

\end{document}